# Knowing When to Move: Evidence Accumulation Models of Human Behavior in Traffic

Floor Bontje[1,4*], Felix van Waveren[2*], Leendert van Maanen[1], Bhargav Nallapu[2], Gustav Markkula[3], Arkady Zgonnikov[2]

[1]Utrecht University, Utrecht, Netherlands; [2]Delft University of Technology, Delft, Netherlands; [3]University of Leeds, Leeds, UK; [4]Netherlands Organisation for Applied Scientific Research, The Hague, Netherlands

*Floor Bontje and Felix van Waveren contributed equally to this work
Corresponding author: Floor Bontje, f.bontje1@uu.nl

**Evidence accumulation models (EAMs) describe decisions as noisy evidence accumulating toward a decision boundary. Although widely studied in laboratory tasks, their application to dynamic, real-world behavior has received less attention. Traffic offers a setting where evidence changes as situations unfold, response timing matters, and decisions occur within ongoing control and interaction. This semi-systematic review synthesizes 28 publications (2014–2026) that model traffic decisions and control actions with EAMs. We organize the literature by modeling level (discrete decisions versus continuous action control) and implementation (stand-alone accumulators versus models embedded in broader perception–action or interaction frameworks). We examine how model components are implemented and evaluated across tasks. Overall, EAMs account for a range of dynamic road-user behaviors. Several comparisons suggest that response timing is better described by accumulating evidence as situations develop than by fixed delays or thresholds on momentary evidence, although accumulation models do not consistently outperform other kinds of models. Whether decreasing boundaries improve the fit differs between studies, and the models offer testable accounts of how decision-relevant information is processed. For cognitive scientists, traffic offers a testbed for studying how evidence from changing environments shapes decision-making. For applied modelers, the findings show how EAM components can be adapted to different traffic tasks, and why the choice of components needs to be evaluated for each task. However, most publications that estimated parameters pooled across participants, and most were not tested beyond the data they were fitted to, highlighting the value of participant-level or hierarchical estimation and of testing on new data.**

# 1. Introduction

In many everyday situations, people make decisions under conditions of uncertainty and continuously evolving information. Such decisions can be conceptualized as a process in which evidence favoring alternative actions is accumulated over time until a response is initiated. Evidence accumulation models (EAMs) have been widely used to study decision-making across cognitive psychology and neuroscience (Forstmann et al., 2016; Ratcliff, 1978; Ratcliff & McKoon, 2008; Ratcliff et al., 2016). These models provide a mechanistic account for choices and response times by assuming that decisions are made by the gradual accumulation of noisy evidence until a boundary is reached (Ratcliff, 1978).

Evidence accumulation models have been extensively studied in controlled laboratory paradigms (Forstmann et al., 2016; Mulder et al., 2014; Ratcliff & McKoon, 2008; Ratcliff et al., 2016). Much of the work concerns simplified, abstract decision-making tasks such as binary perceptual discrimination, lexical decision, and recognition memory. In such tasks, stimuli, responses, and decision endpoints are typically discrete and experimentally constrained. More recent work has begun to examine evidence accumulation in more naturalistic contexts, including air-traffic control, driving, forensic and medical image discrimination, and maritime surveillance (see Boag et al., 2023, for a recent review). Road traffic and road-user behavior represent a particularly relevant applied context within this broader shift, as decisions unfold under time pressure and often involve continuous control and sustained perception–action coupling.

Applying EAMs to human traffic behavior introduces theoretical and computational challenges that go beyond those typically encountered in discrete trial-based tasks. Road users must continuously interpret dynamic environmental cues while selecting and executing actions (Markkula et al., 2018). In such settings, the evidence available to the decision maker may evolve over time, the modeled response may involve action initiation or repeated control adjustments rather than only a discrete choice (Markkula et al., 2018; Pekkanen et al., 2022; Xue et al., 2018), and decision criteria may adapt to situational demands (Liu et al., 2024; Theisen et al., 2024). EAMs in these settings may therefore connect evidence accumulation not only to perceptual input and action execution, but also to the dynamic and interactive context in which the decisions are made.

Over the past decade, this modeling challenge has motivated a growing body of work applying EAMs to traffic behavior across several task paradigms, including pedestrian crossing, emergency braking, vehicle overtaking, and gap acceptance (Mohammad et al., 2024; Pekkanen et al., 2022; Svärd et al., 2021; Zgonnikov et al., 2024a). However, this literature remains fragmented across task paradigms and modeling approaches, making it difficult to obtain a coherent overview of how evidence accumulation models are implemented, parameterized, and evaluated in traffic contexts. A structured synthesis is therefore needed to clarify how EAMs have been applied to human traffic behavior, what modeling assumptions and model evaluation approaches have been used, and what methodological challenges remain.

The present review addresses this gap by synthesizing applications of EAMs to human traffic behavior using a semi-systematic approach. Given the breadth of traffic-related behavior, the search strategy combined an initial exploratory mapping of EAM and traffic-decision literature with a focused database query targeting driver decision-making and maneuver-related traffic behaviors. This was supplemented by citation chaining to identify further work on pedestrians, road-user interaction, and other traffic paradigms. Because this approach does not guarantee exhaustive coverage, we treat the review as semi-systematic rather than systematic. We classify the reviewed models by modeling level (discrete decision-making or continuous action control) and by implementation (stand-alone or embedded in a broader framework), and describe their components, fitting and evaluation. We then synthesize the findings across publications, including where results differ and the limits of the available evidence.

The focus is on human traffic behavior in the form of decisions and control actions used to navigate traffic, such as braking, crossing, merging, overtaking, and steering. This includes drivers and other road users, such as pedestrians. Other traffic-related responses, including responses measured in secondary-task or detection paradigms, may also influence traffic behavior but fall outside our scope when the traffic setting primarily provides the task context or concurrent load rather than the decision or control action being modeled. Research on driver distraction and multitasking is therefore not reviewed here, except where a study models the traffic decision or control action itself. The review is intended for two primary audiences. First, for cognitive scientists, traffic offers a testbed for investigating how evidence from changing environments shapes decisions, and a source of inspiration for new laboratory paradigms. Second, applied modelers in human factors, transportation science and other naturalistic domains may draw on this synthesis when selecting, adapting, and evaluating evidence accumulation models for dynamic traffic tasks. By providing a structured overview of how EAMs are implemented across traffic task paradigms, this review aims to clarify current modeling practices, show where the evidence remains limited, and highlight opportunities for advancing the study of decision-making in dynamic, real-world environments.

# 2. Evidence Accumulation Models and the Drift–Diffusion Framework

This section provides a brief conceptual overview of evidence accumulation models (EAMs), focusing on the core components and the drift–diffusion framework relevant for interpreting the traffic-related modeling work reviewed later in the paper. It summarizes the key concepts needed for the review rather than providing a comprehensive account of EAM theory; readers already familiar with EAMs may wish to skip this section.

## 2.1 Evidence Accumulation Architectures

Evidence accumulation models are a family of computational models that conceptualize decision-making as a process in which noisy evidence is gradually accumulated until a decision boundary is reached (Ratcliff & McKoon, 2008; Ratcliff et al., 2016). The **decision variable** represents the evidence accumulated over time in favor of a response. Its systematic rate and direction of change (i.e., the drift rate**,** hereafter **drift**) is determined by the momentary evidence available to the decision maker, which the accumulation process integrates over time. The starting point (which can encode a response **bias**) can reflect prior knowledge, expectations, or an initial preference toward a particular response, by shifting accumulation closer to the boundary associated with that response. A decision is made once the accumulated evidence reaches the **decision boundary,** which specifies the level of accumulated evidence required for commitment to a response. **Noise** in the accumulation process contributes to variability in both response times and, where alternatives are modeled, choices. In addition, EAMs may account for **non-decision time** to capture processes such as sensory encoding and motor execution that occur outside the decision process itself. These key components of EAMs are also summarized in Table 1.

EAMs differ in several respects. A first distinction concerns the number of response alternatives a model represents. In two-boundary models, also called two-choice models, the decision variable moves between two boundaries, each standing for one identifiable response, and the model predicts both which response occurs and when (Ratcliff, 1978, 2015; Ratcliff & McKoon, 2008). In single-boundary models, also called one-boundary or one-choice models, accumulation runs to a single boundary and the model predicts only the timing of one action, since there is no alternative response for the accumulation to be directed away from (Anders et al., 2016; Ratcliff & Strayer, 2014; Ratcliff & Van Dongen, 2011). Models with more than two response alternatives also exist and are typically implemented as races between separate accumulators (see below). Figure 1 illustrates how evidence accumulation determines response selection and timing in two-boundary and single-boundary models.

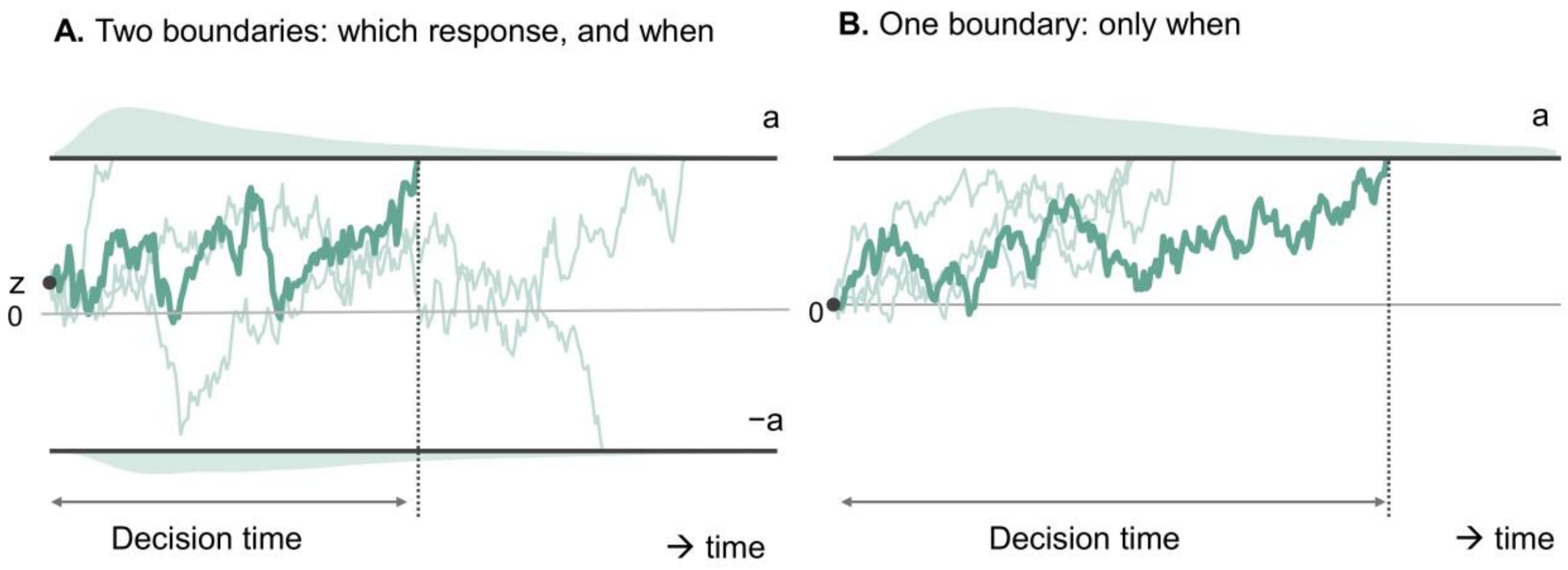

***Fig. 1 Two-boundary and single-boundary evidence accumulation models.*** *Each panel shows four simulated trials, one highlighted in bold; within-trial noise makes trials differ in decision time and, in panel A, in the response. The dotted line marks the decision time of the highlighted trial, and the shaded areas show the decision-time distributions. (A) Two-boundary model: evidence accumulates from starting point z, here shifted toward a (a bias toward that response), until it reaches a or −a, each representing one response (e.g., cross or wait); the drift is the mean rate and direction of accumulation. (B) Single-boundary model: evidence accumulates from 0 to boundary a, representing initiation of one response (e.g., braking), so the model predicts when, not which, response occurs.*

Beyond the number of response alternatives, EAMs also differ in how evidence for those alternatives is represented. In relative evidence models, with the drift–diffusion model (DDM) serving as a common reference example (Section 2.2), a single decision variable integrates the difference in evidence between alternatives. The model is also known as the diffusion decision model (Ratcliff & McKoon, 2008; Ratcliff et al., 2016); we use drift–diffusion model throughout. In absolute multi-accumulator models, or race models, evidence for each response option is accumulated independently by separate accumulators that race toward their decision boundaries (Smith & Ratcliff, 2004); the linear ballistic accumulator is a widely used example (Brown & Heathcote, 2008). Further evidence accumulation model variants exist but are beyond the scope of this section (for overviews, see Bogacz et al., 2006; Smith & Ratcliff, 2004).

*Table 1* *Key components of evidence accumulation models.*

| Component | Description | Role in the accumulation process |
|---|---|---|
| **Decision variable** | Current state of the accumulated evidence | Tracks the accumulation process over time and determines when a decision boundary is reached |
| **Drift** | Systematic rate and direction of change of the decision variable; may be constant or vary over time | Reflects the strength and direction of the evidence driving the change of the decision variable |
| **Starting point** | Initial state of the decision variable relative to the decision boundary or boundaries | Can reflect an initial preference toward a response by placing the decision variable closer to its associated boundary |
| **Decision boundary** | Level of accumulated evidence required for response commitment (fixed or decreasing) | Represents response caution; a decreasing boundary can represent urgency |
| **Noise** | Stochastic variability in the accumulation process | Captures within-trial fluctuations in evidence accumulation, contributing to variability in response times and, where alternatives are modeled, choices, even under similar stimulus conditions |
| **Non-decision time** | Time attributed to processes outside the evidence-accumulation process, such as sensory encoding and motor execution | Accounts for perceptual and motor delays not related to decision formation |

## 2.2 The Drift–Diffusion Model

The DDM (illustrated in Figure 1) is one of the most widely used relative evidence models and the basis for many extensions. Nearly all reviewed publications use a DDM or a variation on the DDM (Section 4), so this section provides a closer overview of this model type. We consider both its two-boundary form for modeling alternative responses and its single-boundary form for modeling the timing of a single action.

In one common formulation, the DDM describes a single decision variable, D(t), that accumulates noisy evidence from a starting point z until it reaches one of two decision boundaries, $a$ a or $-a$ , at which a response is initiated. Formally, the drift–diffusion process is expressed as:

$$dD = v\,dt + \sigma\,dW(t), (1)$$

where $dD$ is the change in the decision variable over the time increment $dt$, $v$ specifies the drift, the rate at which the momentary evidence available to the decision maker is integrated, and $\sigma\,dW(t)$ represents within-trial noise, the increment of a Wiener process scaled by $\sigma$ (Ratcliff, 1978; Ratcliff & McKoon, 2008). Within-trial noise is distinct from across-trial variability, in which the drift, the starting point or the non-decision time varies from one trial to the next; the full drift–diffusion model includes parameters for such variability (Ratcliff & McKoon, 2008). Here the drift is constant within a trial; extensions in which $v$ varies over time within a trial, as the situation unfolds, are discussed below and are of particular relevance to traffic applications. Variation in the drift, starting point, and boundary separation allows the model to capture systematic effects of stimulus quality, response bias, and response caution (Bogacz et al., 2006; Forstmann et al., 2016). In particular, wider boundary separation generally produces slower but more accurate responses, whereas narrower separation produces faster but more error-prone responses, a pattern known as the speed–accuracy trade-off.

Various extensions of the standard DDM have been proposed to capture more complex decision dynamics (Forstmann et al., 2016; Ratcliff & McKoon, 2008; Ratcliff et al., 2016). In the context of this review, the most relevant variations concern how evidence evolves over time and how it is integrated into the decision process. Time-varying drift allows the evidence input to change as a situation unfolds, which is especially relevant in dynamic environments and is specified in nearly all reviewed publications (Section 4.3.1). Leaky accumulation modifies how past evidence is retained by allowing previously accumulated information to decay over time; it is part of at least seven of the reviewed models (Durrani & Lee, 2023, 2024; Engström et al., 2024; Giles et al., 2019; Markkula et al., 2023; Pekkanen et al., 2022; Svärd et al., 2021). Urgency mechanisms make decisions easier as time passes, for example by multiplying the evidence by a signal that grows over time (Cisek et al., 2009); a decreasing boundary, described below, has a similar effect. Finally, in interacting-accumulator models, separate accumulators for the alternative responses inhibit one another (Usher & McClelland, 2001); such architectures are briefly considered as directions for future work in Section 6.1.

Beyond accumulation dynamics, the specification of decision boundaries represents a second major source of model variation (Section 4.3.2). Fixed boundaries imply stable decision criteria, whereas decreasing boundaries, denoted $a(t)$, reduce the evidence required as time elapses, which is commonly interpreted as increasing urgency. Time pressure is a property of the task or the environment; a decreasing boundary is one way of representing the urgency that time pressure may produce, rather than a direct measure of it.

A third source of variation is the starting point. A starting point shifted toward one boundary means that less evidence is needed to reach it, making that response faster and more likely; in a gap-acceptance task, for example, it reflects a prior tendency to accept the gap regardless of the evidence from the approaching vehicle. In two-boundary models, asymmetric boundaries can likewise favor one response over another. In single-boundary models, a shifted starting point cannot be distinguished from a change in the boundary, as the distance between the two influences the response time; prior expectation could therefore be reflected in this distance but also in other parameters, such as the drift (Ratcliff, 2015; Ratcliff & Strayer, 2014). Together, boundary separation, boundary dynamics and starting point bias shape the speed–accuracy trade-off, response tendencies and response time distributions without altering the evidence input (Forstmann et al., 2016; Ratcliff & McKoon, 2008).

To conclude, the drift, decision boundary, starting point, noise and non-decision time provide a flexible framework for modeling decision-making as evidence accumulation. The next section describes how publications applying these models to traffic were identified and reviewed.

# 3. Method

This review used a two-phase literature search: an exploratory phase followed by a focused search. The exploratory phase aimed to provide initial context and theoretical grounding through an iterative literature exploration based on foundational publications and successive keyword-driven database searches (Supplementary Materials, exploration phase and Table S1). This phase was intended to build theoretical grounding and to develop the eligibility criteria for the focused review, rather than to assemble the set of reviewed publications itself. The broad terms used at this stage, including *modeling methods*, therefore deliberately extended beyond traffic applications in order to capture the general evidence accumulation literature on which the criteria were based. The result was a set of 420 records, reduced to 209 after removing duplicates, of which 71 were retained on title and 61 after screening abstracts for relevance to traffic decisions, decision modeling, modeling methods, or combinations of two or more of these topics. Based on a high-level examination of these 61 publications, a set of specific inclusion and exclusion criteria was defined for the focused review presented in this paper:

**Inclusion criteria:**

1. An evidence accumulation model had to be explicitly implemented and identifiable in the publication, meaning that the accumulated quantity and the criterion for response initiation were both stated, through equations, a published specification or an explicit schematic, and that the account was evaluated against behavioral data or a competing account of the same behavior. Formal parameter estimation was not required. The evaluation had to concern the accumulation process itself, that is, its predictions of response timing or choice. This included qualitative comparisons without fitting (e.g., Markkula et al., 2016, who apply the accumulation account of Markkula, 2014, to naturalistic braking) and evaluations of a simplified form of the account (Engström et al., 2024), but not models in which accumulation was added only to illustrate variability in otherwise deterministic model behavior (Lin et al., 2022; Supplementary Table S7).
2. The modeled behavior had to be a decision or control action by which a human road user (driver, pedestrian or cyclist) navigates traffic, such as braking, crossing, merging, overtaking or steering.
3. The model had to relate the modeled decision or control action to the traffic situation: the accumulated evidence, or parameters estimated per condition, had to vary with properties of that situation, such as looming, time-to-arrival, a gap or a control error. Publications that also examined task or participant factors, such as instructions, age or a concurrent task, were included when the model was also linked to the traffic situation, and excluded when it was related to such factors only.
4. Peer-reviewed journal articles and full conference papers, published in English from 2014 onward, as this corresponds to the earliest identified application of EAMs to traffic decision modeling during the exploratory phase (Markkula, 2014).

**Exclusion criteria:**

1. Publications applying evidence accumulation models outside traffic contexts or to secondary tasks performed while driving, including publications in which driving only provides the context or workload for responses to a detection signal. Where a publication reports both an eligible traffic application and such a secondary task, only the traffic application is coded.
2. Publications centered on automated driving takeover or human–automation control sharing or transfer decisions, as the review focused specifically on human-controlled traffic decisions.
3. Publications discussing driver behavior or traffic decisions without explicit evidence accumulation modeling.
4. Duplicate publications.

Using these criteria to guide term selection, we constructed the following Scopus search query:

```
ALL (((evidence AND accumulation) OR (drift AND diffusion) OR (lba) OR
(linear AND ballistic AND accumulator)) AND (driver AND decision) AND
(merging OR (gap AND acceptance) OR (steering) OR (lane AND change) OR
overtaking OR braking OR passing OR (stop AND go) OR (obstacle AND
```

```
avoidance) OR (lead AND following)) AND model AND cognitive AND traffic AND
NOT takeover) AND (LIMIT-TO (SUBJAREA, "ENGI" ))
```

Besides drift–diffusion terms, the query included the linear ballistic accumulator, so that this multi-accumulator model could be retrieved if it had been applied in traffic settings. The query was last run on 13 October 2026, and the focused-search counts reported below refer to that run. The focused search used Scopus only, and the query was applied to all fields of the records (field code ALL). It was limited to the Scopus engineering subject area, to focus on work approaching evidence accumulation from an application perspective; this filter classifies publication venues rather than study content and was not applied to publications found through citation chaining. The run returned 105 records. After their titles and abstracts were screened against the inclusion and exclusion criteria, 17 full texts were sought; two could not be retrieved, and of the 15 assessed, 14 were included and one was excluded (Supplementary Table S7).

Backward citation chaining (checking the reference lists of retrieved records) was applied in the first run, to the records retained after title and abstract screening; forward chaining was not performed. As the last run identified no eligible publications beyond those already included, chaining was not repeated. Screening also identified several diffusion-model publications by Ratcliff and colleagues that model responses in a driving simulator but relate the model parameters to task or participant conditions rather than to the traffic situation (inclusion criterion 3; Ratcliff, 2015; Ratcliff & Strayer, 2014; Ratcliff & Vanunu, 2022; Vanunu & Ratcliff, 2023). These and the evidence accumulation publications on driver distraction (exclusion criterion 1; Castro et al., 2019, 2023; Tillman et al., 2017) were excluded because of the behavior and evidence being modeled, not their methodological quality. Supplementary Table S7 lists them with the reason, and Section 2.2 draws on the excluded diffusion-model publications for the single-boundary form of the model.

Supplementary Figure S1 shows the search and screening process for both phases, including how the 28 included publications were identified. The coding categories used in the extraction are defined in the note to Table 2 and in the codebook (Supplementary Table S2). The full extraction sheet is provided in Supplementary Tables S3 and S4. In the last run, the query was also run without the subject-area filter, which returned 424 records. The 319 records that the filter had removed contained four publications already found through citation chaining, one of the excluded diffusion-model publications (Ratcliff & Vanunu, 2022), and no further eligible publications. At the time of the last run, all 28 included publications had appeared as versions of record, and they are cited as such.

For the first run of the search, screening against the eligibility criteria was carried out by one author (FvW), including review of the reference lists of the retrieved records; later runs were screened by another author (FB). The resulting set of included publications was subsequently checked by the remaining authors, who are familiar with the relevant literature. Twenty-three of the 28 included publications have at least one author of this review as a co-author; they were judged against the same eligibility criteria and coded with the same definitions as the other publications. Screening was therefore sequential rather than an independent duplicate assessment, and inter-rater agreement was not quantified. Uncertain coding decisions were discussed among the authors until agreement was reached. FvW, BN and LvM are not authors of any reviewed publication, and FB is an author of one (Bontje & Zgonnikov, 2024).

The unit of analysis in this research was the publication; publications that analyze the same dataset or extend the same model are linked in Supplementary Tables S5 and S6. Accordingly, counts in this review refer to publications; a publication may report several studies, that is, experiments or analyses, and is then counted once. Each publication was analyzed to extract information on the decision maker and traffic task; the modeled response, that is, the behavioral outcome addressed by the model; the perceptual or cognitive quantity the drift is a function of; the specification of the drift, the decision boundary and the starting point; the level at which parameters were estimated (i.e., per participant or for data pooled across participants); the data source and sample size; the models the accumulation model was compared with; and the model evaluation evidence reported. These categories follow the model components described in Section 2 (the evidence input and drift, the decision boundary and starting point), complemented by the study characteristics needed to evaluate an application (task, modeled response, parameter estimation, data, model comparison and model evaluation); the note to Table 2 and the Supplementary Materials define each category. The synthesis involved grouping publications by these themes and comparing the reported model specifications, evaluation approaches, and findings across publications. The identified trends informed the structure of this review's sections. No formal quality or risk-of-bias assessment was conducted; instead, Table 2 summarizes the model evaluation evidence reported in each publication.

The original literature selection was conducted manually. A large language model (Claude, Anthropic) assisted with preliminary title and abstract screening in the last run, and with checking table entries and extracted data against the source publications. One author (FB) verified every output and made the final screening and coding decisions in the last run. All authors reviewed and approved the final manuscript and take full responsibility for its content.

# 4. Overview of the Reviewed Models

In this section, we first classify the reviewed publications by modeling level, distinguishing discrete decision-making from continuous action control (Section 4.1), and by model implementation, distinguishing stand-alone models from implementations embedded in broader perception–action or interaction frameworks (Section 4.2). We then review how the core EAM components are specified (Section 4.3) and which data, fitting procedures and model evaluation the publications use (Section 4.4). Section 5 then synthesizes what the reviewed publications show across studies.

Table 2 summarizes, for each publication, the traffic task, the implementation of the EAM, the modeled response, the specification of the model components, the estimation level, the models compared, the reported model evaluation evidence and the key reported finding; the note below the table defines each category, including the types of model evaluation evidence. Sample sizes, data sources and detailed coding notes are provided in Supplementary Tables S3 and S4.

***Table 2 Overview of 28 publications applying evidence accumulation models of human behavior in traffic: model specification, models compared, model evaluation evidence and key reported findings***. *Publications are grouped by modeling level and, within each level, by traffic task; within each task group they are ordered by publication year, with the originating publication of a model placed before those that extend it.*

| Publication | Implementation | Modeled response | Evidence input | Drift | Decision boundary | Starting point | Estimation | Models compared | Model evaluation evidence | Key reported findings |
|---|---|---|---|---|---|---|---|---|---|---|
| **Discrete decision-making — the model represents a single decision with a defined outcome** | | | | | | | | | | |
| **Pedestrian crossing** | | | | | | | | | | |
| Giles et al. (2019) | Stand-alone | Crossing initiation time | Time-to-arrival and its rate of change | Time-varying | Fixed | Not shifted | Pooled | Alternative accumulator specifications | Model selection | A single accumulator with a time-varying, bounded drift reproduced crossing times nearly as well as a two-stage model. |
| Pekkanen et al. (2022) *yielding vehicle* | Stand-alone | Choice; crossing initiation time | Time-to-arrival, distance, rate of change of time-to-arrival, eHMI | Time-varying | Fixed | Not shifted | Pooled | Alternative accumulator specifications | Model selection; cross-dataset transfer (two parameters re-estimated) | Distance and the rate of change of time-to-arrival added to time-to-arrival, and an eHMI signal acted as a fixed evidence boost. |
| Theisen et al. (2024) | Stand-alone | Choice; response time (crossing and waiting) | Time-to-arrival weighted by vehicle speed (3 of 7 models); none, drift free per condition (4 of 7) | Time-varying (2 of 7 models); condition-dependent (5 of 7) | Decreasing (in time-to-arrival; 1 of 7 models, the others fixed) | Shifted (2 of 7 models, free per condition) | Pooled | Alternative accumulator specifications | Model selection | Estimated per condition, drift rose with time-to-arrival and the boundary was U-shaped; the preferred model depended on the criterion. |
| Ma et al. (2025) | Stand-alone | Choice; response time | Time-to-arrival, or anticipated utility of crossing | Time-varying | Decreasing (in time-to-arrival; fixed also tested) | Not shifted | Individual | Alternative accumulator specifications | Model selection; correspondence with an independent measure | A decreasing boundary provided a better fit than a fixed one, and EEG results were more consistent with the utility-based drift than with the time-to-arrival drift. |
| Tian et al. (2025) *yielding vehicle* | Stand-alone | Choice; crossing initiation time | Time-to-arrival, distance, rate of change of time-to-arrival | Time-varying | Fixed | Not shifted | Pooled (not refitted) | Against a discrete-choice model with a perceptual threshold, and a neural network | Model selection | Reused without refitting, the accumulation model matched the data in fewer conditions than a perceptual-threshold choice model. |
| Ma et al. (2026) *cognitive load* | Stand-alone | Choice; response time | Time-to-arrival or utility, scaled by cognitive load | Time-varying | Decreasing (in time-to-arrival) | Not shifted | Individual | Alternative accumulator specifications | Model selection; correspondence with an independent measure | Cognitive load was best modeled as a weaker response to the evidence, which the EEG supported. |
| **Gap acceptance: left turn** | | | | | | | | | | |
| Zgonnikov et al. (2024a) | Stand-alone | Choice; response time (go decisions only) | Time-to-arrival and distance | Time-varying | Decreasing (in time-to-arrival; fixed also tested) | Not shifted | Individual | Alternative accumulator specifications | Model selection; cross-condition generalization | Choices were fitted with a fixed boundary, but only a model with a time-varying drift and a decreasing boundary captured the response-time pattern. |
| Zgonnikov et al. (2024b) *accelerating* | Stand-alone | Choice; response time | Time-to-arrival and distance | Time-varying | Decreasing (in time-to-arrival; | Shifted | Pooled | Alternative accumulator specifications | Descriptive fit; model selection; cross-condition generalization | A starting point bias toward going improved the fit, a decreasing boundary did not improve the fit, and withheld maneuvers were largely predicted. |

| Publication | Implementation | Modeled response | Evidence input | Drift | Decision boundary | Starting point | Estimation | Models compared | Model evaluation evidence | Key reported findings |
|---|---|---|---|---|---|---|---|---|---|---|
| *or decelerating vehicle* | | | (acceleration tested and rejected) | | tested, fixed preferred) | | | | | |
| Bontje & Zgonnikov (2024) *confidence rating* | Stand-alone | Choice; response time; confidence rating (second fitting stage) | Time-to-arrival and distance | Time-varying | Decreasing (in the generalized gap) | Not shifted | Pooled | Alternative accumulator specifications | Model selection; correspondence with an independent measure (decision parameters fitted without the confidence ratings) | Choices and response times did not separate a race from a relative accumulator; confidence favored a single accumulator that kept accumulating after the decision. |
| **Gap acceptance: overtaking** | | | | | | | | | | |
| Mohammad et al. (2023) | Stand-alone | Choice; response time | Time-to-arrival and distance (plus initial speed in four of eight variants) | Time-varying | Decreasing (in the generalized gap) | Shifted (speed-dependent) | Pooled | Alternative accumulator specifications | Model selection | Only models with a starting point that shifted toward overtaking at higher speed reproduced all response-time patterns. |
| Mohammad et al. (2024) *decelerating vehicle* | Stand-alone | Choice; response time | Time-to-arrival and distance | Time-varying | Decreasing (tested, fixed preferred) | Shifted (speed-dependent) | Pooled | Alternative accumulator specifications | Model selection | Refitted to new data, the speed-dependent starting point again improved the fit and a decreasing boundary did not improve the fit. |
| **Yellow-light decision** | | | | | | | | | | |
| Liu et al. (2024) | Stand-alone | Choice; response time | None (drift free per condition) | Condition-dependent | Fixed | Not shifted | Individual | None | Descriptive fit; cross-condition generalization | Drift shifted from going to stopping as the time to the stop line grew; fitted parameters predicted a withheld condition. |
| **Brake timing** | | | | | | | | | | |
| Markkula (2014) | Stand-alone | Looming at detection; looming at brake onset | Looming (angular expansion rate; inverse tau) | Time-varying | Fixed | Not shifted | Pooled | None | Descriptive fit | Accumulated looming reproduced detection thresholds and braking behind decelerating vehicles but braked too early behind stationary ones. |
| Markkula et al. (2016) *rear-end conflicts* | Embedded | Brake onset time; jerk; maximum deceleration | Looming (proposed, not fitted) | Proposed, not fitted | Not fitted | Not applicable | None | None (no accumulator fitted) | Cross-dataset transfer (of the descriptive model); qualitative comparison | Brake onset followed looming rather than a fixed delay; the accumulation account of Markkula (2014) is compared with this pattern qualitatively, without fitting. |
| Svärd et al. (2017) *pre-crash* | Embedded | Brake onset time; ramp-up (simulated) | Prediction error on looming | Time-varying | Fixed | Not shifted | None | None | Descriptive fit | A hand-tuned accumulation model reproduced faster, steeper braking at higher looming, but braked later than drivers in near-crashes. |
| Xue et al. (2018) | Stand-alone | Brake onset time | Looming (angular expansion rate; inverse tau); brake lights | Time-varying | Fixed | Not shifted (cue enters the drift) | Pooled | Alternative accumulator specifications; against a perceptual-threshold model; against a statistical model | Model selection; held-out prediction | Accumulator models fitted brake response times better than looming-threshold models, and the brake lights influenced the fit; a multilevel regression fitted to all data outperformed the accumulator model. |
| Svärd et al. (2021) *off-road glances* | Embedded | Brake onset time; jerk | Prediction error on looming, downweighted in off-road glances | Time-varying | Fixed | Not shifted | Pooled | Alternative accumulator specifications | Descriptive fit; model selection | Letting looming accumulate at reduced weight during off-road glances improved the fit to real crashes and near-crashes. |

| Publication | Implementation | Modeled response | Evidence input | Drift | Decision boundary | Starting point | Estimation | Models compared | Model evaluation evidence | Key reported findings |
|---|---|---|---|---|---|---|---|---|---|---|
| Durrani & Lee (2023) *car following* | Stand-alone | Response time (accelerator release on approach; brake onset), fitted separately | Looming (angular expansion rate; inverse tau); brake lights | Time-varying | Fixed | Not shifted (cue enters the drift) | Individual (per drive) | Alternative accumulator specifications; against a perceptual-threshold car-following model | Model selection; held-out prediction | Accumulation predicted held-out response times better than the Wiedemann thresholds; brake-light evidence was not assigned a measurable contribution. |
| **Collision avoidance** | | | | | | | | | | |
| Engström et al. (2024) | Embedded | Response time (evasive maneuver onset) | Surprise relative to a learned prior | Time-varying (computational model only) | Fixed | Not shifted (a learned prior shapes the evidence) | None (accumulator fitted only to model-generated response times) | Against a fixed-reaction-time model (regression model only) | Descriptive fit; cross-dataset transfer (both of the regression model) | Response time grew with the build-up of surprising evidence, which a fixed reaction time misses; the accumulator was fitted only to response times generated by the regression model. |
| Li et al. (2024) *emergency* | Stand-alone | Choice (brake or steer); response time | Initial speed, and time headway and distance to the lead and to the conflicting vehicle | Time-varying | Decreasing (in time headway and distance) | Shifted (speed-dependent) | Pooled | Alternative accumulator specifications | Descriptive fit; model selection | Only models with a speed-dependent starting point and drift from both vehicles reproduced faster braking at higher speeds. |
| Huang et al. (2025) *emergency* | Embedded | Choice (brake or steer); response time | Initial speed, and time headway and distance to the surrounding and conflicting vehicles | Time-varying | Decreasing (in time headway and distance) | Shifted (speed-dependent) | Pooled (one parameter set per scenario) | Against conventional car-following and lane-change models | Descriptive fit; model selection | Steering choices were captured after lead-vehicle braking but not after cut-ins; the model avoided simulated collisions. |
| **Continuous action control — accumulation to a boundary recurs within an ongoing control loop** | | | | | | | | | | |
| **Longitudinal control (car following)** | | | | | | | | | | |
| Durrani et al. (2021) | Stand-alone | Looming and accumulated looming at response | Looming (inverse tau) | Time-varying (no gain, noise or leak) | Fixed (not fitted) | Not shifted | None | Against perceptual-threshold and reaction-time accounts (hypotheses, nothing fitted) | Qualitative comparison | Fixed thresholds and reaction times did not explain response timing, whereas accumulated inverse tau did, more consistently when approaching than when braking. |
| Durrani & Lee (2024) | Embedded | Distance to lead vehicle (speed, acceleration and jerk compared, not fitted) | Prediction error on the control error | Time-varying | Fixed (asymmetric) | Not shifted (asymmetric boundaries) | Individual (per drive); pooled (field data) | Alternative accumulator specifications; against conventional car-following models, one of them perceptual-threshold based | Model selection; held-out prediction (field data only) | Accumulator-triggered control reproduced intermittent adjustments but had the lowest pointwise error only when approaching a stopped vehicle. |
| **Lateral control (steering)** | | | | | | | | | | |
| Markkula et al. (2018) | Embedded | Steering onset time; amplitude; interval | Prediction error on the perceptual control error | Time-varying | Fixed | Not shifted | Mixed (individual | Against continuous control and a threshold-plus-refractory model | Model selection | Steering was better described as discrete adjustments than as continuous control, with timing that tentatively favored accumulation over an error threshold. |

| Publication | Implementation | Modeled response | Evidence input | Drift | Decision boundary | Starting point | Estimation | Models compared | Model evaluation evidence | Key reported findings |
|---|---|---|---|---|---|---|---|---|---|---|
| | | | | | | | and pooled) | | | |
| Goodridge et al. (2023) | Stand-alone | Steering onset time; lateral position error; steering rate | Perceptual control error from visual angle | Time-varying (no gain, noise or leak) | Fixed (not fitted) | Not shifted | None (not fitted) | Against a perceptual-threshold model | Qualitative comparison | Steering onset depended on how fast the error grew, as accumulation predicts and a fixed threshold does not. |
| **Collision avoidance** | | | | | | | | | | |
| Schumann et al. (2026) | Embedded | Brake and steering onset time; choice; collisions; deceleration | Surprise: residual pragmatic value of the current policy | Time-varying | Fixed | Not shifted (a norm-conditioned prior) | Pooled (manual tuning) | Ablation of components | Model selection; cross-condition generalization; cross-dataset transfer | Removing accumulation made responses substantially faster than observed; the full model largely matched new intersection data without retuning. |
| **Road-user interaction** | | | | | | | | | | |
| Markkula et al. (2023) | Embedded | Five interaction phenomena (speed and deceleration metrics); crossing initiation time; who passes first | Action value under noisy perception, weighted by the other agent's likely behavior | Time-varying | Fixed (relative) | Not shifted (a Bayesian prior over the other agent) | None (grid search) | Ablation of components | Model selection; cross-dataset transfer | Only variants with Bayesian perception and accumulation reproduced the two-peaked crossing times. |
| Schumann et al. (2023) *gap acceptance* | Embedded | Choice; time at conflict point | Action value under noisy perception, weighted by the other agent's likely behavior | Time-varying | Fixed (relative) | Not shifted | Pooled | Alternative model and fitting configurations; against a trajectory prediction model and a logistic regression | Model selection; held-out prediction | The model matched or performed at least as well as data-driven models in most comparisons; what accumulation itself adds was not tested. |

*Note. Decision maker: the pedestrian in the pedestrian crossing group, the driver and the pedestrian in Markkula et al. (2023), and the driver in all other publications. Variants of a task group are named in italics beneath the publication. Time-to-arrival, time for the relevant vehicle to reach the conflict point (in some models its optical estimate, tau); time headway, distance divided by speed; generalized gap, time-to-arrival plus weighted distance; looming, optical expansion (variable in parentheses). Modeled response, the behavioral outcome addressed by the model (which was not necessarily fitted). Evidence input, the quantity the drift is a function of. Drift: time-varying, changes with the input during accumulation; condition-dependent, constant within a trial but different between conditions. Decision boundary: fixed, constant during accumulation; decreasing, decreases during accumulation with the named variable, coded when at least one model of the publication has such a boundary. Implementation: stand-alone, the accumulation model is applied directly to the behavioral data; embedded, accumulation is one component of a broader perception–action or interaction framework. Starting point: shifted, the initial state of the decision variable is closer to one of two boundaries, favoring that response; not shifted, no such shift is modeled, which includes all single-boundary models; other sources of bias are noted in parentheses. Estimation: pooled, one parameter set for all participants together; individual, estimated per participant or, where each drive is fitted separately, per drive; mixed, some parameters individually and some at group level; none, no accumulator parameters estimated from the behavioral data analyzed. Not fitted, set by hand or proposed without estimation. Models compared: alternative accumulator specifications, variants of the accumulation model compared on the same data; against a named model of another kind; ablation of components, mechanisms removed one at a time. Model evaluation evidence: descriptive fit, comparison with the data used for fitting; model selection, fitted model specifications compared on the same data; qualitative comparison, predicted and observed patterns contrasted without fitting both alternatives; held-out prediction, evaluation on observations withheld from fitting; cross-condition generalization, prediction of an experimental condition withheld from fitting; cross-dataset transfer, evaluation on an independent dataset; correspondence with an independent measure, an estimated quantity compared with a measure not used in fitting. A publication may fall into more than one category, and the categories are not a quality ranking. Where the evaluated model is not the accumulation model, the cell names it. Key reported findings, the main reported result in one line. Sample sizes, data sources, full parameterization and coding notes are given in Supplementary Tables S3 and S4.*

## 4.1 Modeling Level

The reviewed literature can be categorized according to the modeling level used: whether EAMs are used to model behavior in a discrete decision-making setup or a continuous action control setup (see banner rows of Table 2). Figure 2 illustrates the distinction between these two modeling levels and indicates which of the reviewed traffic tasks were modeled at each level.

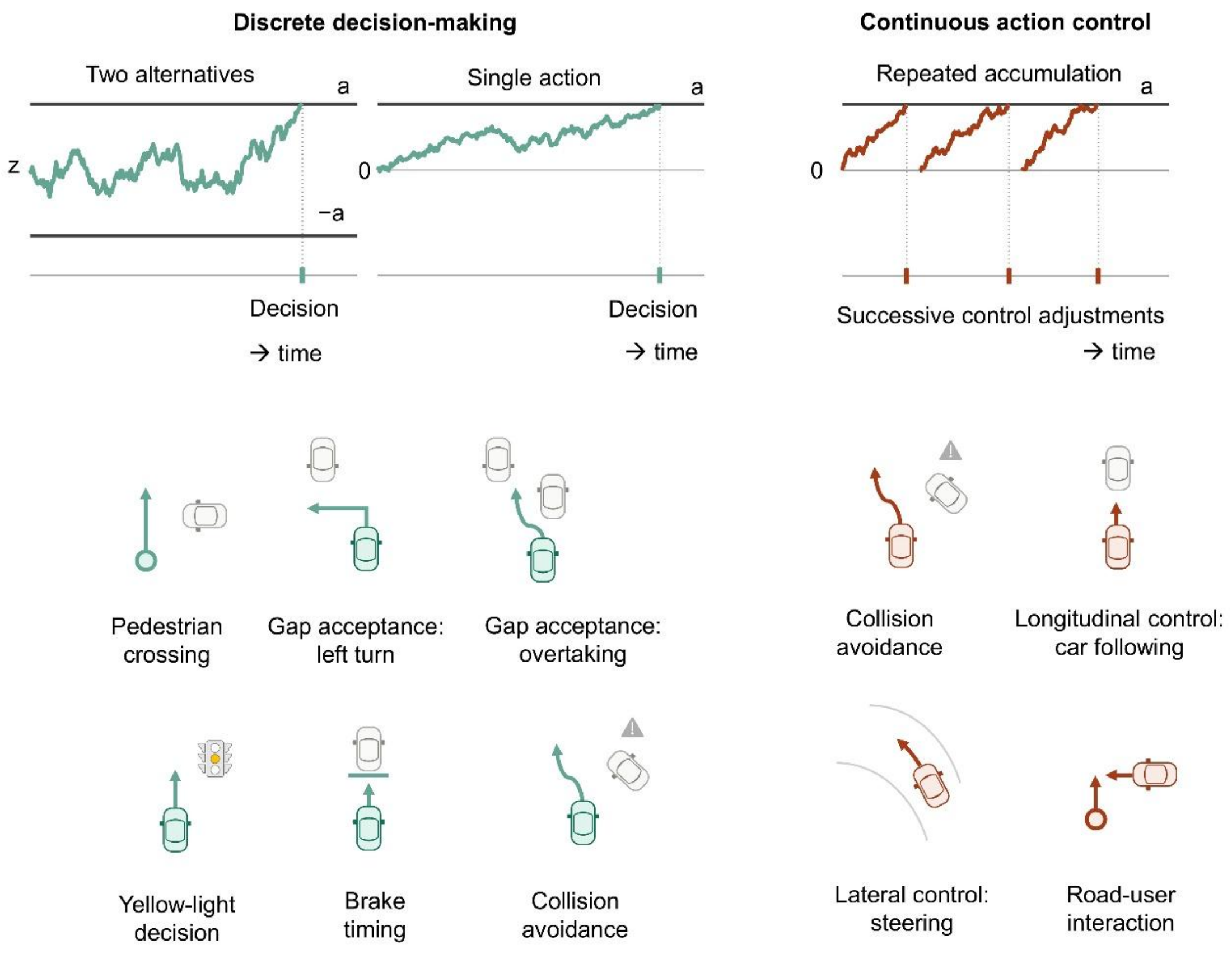

***Fig. 2 The two modeling levels and the traffic tasks they are applied to.*** *Top: accumulation to a boundary as a choice between two alternatives (left) or as the timing of a single action (middle), both modeled as discrete decisions, and single-boundary accumulation repeated within an ongoing control loop (right), modeled as continuous action control. a and −a are boundaries, z and 0 starting points; tick marks show when a response is initiated. Bottom: the traffic tasks of the reviewed publications by modeling level (green, discrete decision-making; rust, continuous action control); decision makers in color, other road users in gray.*

At the discrete decision-making modeling level, tasks involve a single, discrete traffic decision or maneuver that interrupts ongoing behavior. These tasks are characterized by a clear commitment point and a well-defined decision outcome, that is, a point at which the modeled response either occurs or does not, whether that response is a choice between alternatives or the initiation of a single action. From a traffic-engineering perspective, the reviewed publications in this category concentrate on discrete conflict scenarios, particularly intersections and crossing situations. Examples include left-turn crossing maneuvers, overtaking maneuvers, pedestrian crossing decisions, yellow-light decisions, and isolated emergency maneuver responses such as braking or evasive steering (Giles et al., 2019; Li et al., 2024; Liu et al., 2024; Mohammad et al., 2023; Pekkanen et al., 2022; Tian et al., 2025; Zgonnikov et al., 2024a). Eleven of the 21 publications at this modeling level concern a gap-acceptance task (pedestrian crossing, left turn and overtaking), in that road users must decide whether the available time or space is sufficient to execute an action safely. Six publications concern brake timing and focus on when a braking response is initiated (Durrani & Lee, 2023; Markkula, 2014; Markkula et al., 2016; Svärd et al., 2017, 2021; Xue et al., 2018), whereas the three collision-avoidance publications at this level model the onset of an evasive action and, in two of them, the choice between braking and steering (Engström et al., 2024; Huang et al., 2025; Li et al., 2024). Publications on these tasks are classified at the discrete decision-making level when the model primarily predicts action initiation or action selection, rather than sustained control dynamics.

In contrast, publications at the continuous action control modeling level address sustained control tasks, involving ongoing sensorimotor regulation through repeated adjustments. Notably, in most of these models the underlying mechanism remains discrete: accumulation proceeds to a bound, which triggers an adjustment, after which accumulation restarts (Markkula et al., 2018). What distinguishes this modeling level is therefore not the absence of a boundary, but the repetition of accumulation-to-bound events within an ongoing control loop. Examples include longitudinal control during car following (Durrani & Lee, 2024), lateral control in lane keeping (Markkula et al., 2018) and in steering toward curves (Goodridge et al., 2023), and continuous collision avoidance (Schumann et al., 2026). The level refers to this behavioral context, even when the accumulation component accounts only for the timing of individual adjustments. From a traffic-engineering perspective, these publications are more often concerned with road-segment control problems than with discrete conflict points. The reviewed steering publications focus on individual steering corrections (Goodridge et al., 2023; Markkula et al., 2018), because accumulation specifies when an adjustment is triggered, while its magnitude is determined by other components of the control framework.

Some driving behaviors may appear in both modeling level categories. Collision avoidance is the clearest case in the reviewed set, and Figure 2 marks it at both levels. The potential collisions are similar at both levels: a lead vehicle that slows or brakes, or a vehicle that cuts in, with a static obstacle added in Huang et al. (2025) and an intersection in Schumann et al. (2026). The levels differ in whether the model predicts when an evasive action begins (discrete decision-making; Engström et al., 2024; Huang et al., 2025; Li et al., 2024) or how braking and steering are controlled throughout the conflict (continuous action control; Schumann et al., 2026). In this review, sustained longitudinal and lateral control were categorized at the continuous-control level, whereas isolated braking or evasive-steering onsets were categorized at the discrete-decision level. This illustrates how the distinction between discrete decision-making and continuous action control reflects differences in modeling abstraction rather than exclusive behavioral categories. For the same reason, Durrani et al. (2021) and Durrani and Lee (2023), which analyze the same car-following dataset, are coded at different levels: the former also examines how deceleration is regulated after the response, whereas the latter models a single response per drive to the lead vehicle. The road-user interaction models are coded at the continuous level because they simulate the trajectories of the interacting road users, from which choices such as who passes first are derived (Markkula et al., 2023; Schumann et al., 2023).

Having classified the tasks addressed in the literature, and the level at which they are modeled, the following subsections examine the reviewed publications along four further aspects: model implementation (stand-alone versus embedded), the specification of EAM components, data sources, and model evaluation.

## 4.2 Model Implementation: Stand-Alone and Embedded EAMs

The reviewed publications also differ in how EAMs are implemented: as stand-alone decision models or as embedded components within broader perception–action or interaction frameworks (Implementation column of Table 2).

The choice between a stand-alone and an embedded implementation is closely related to the modeling level at which the traffic decision task is being modeled. Stand-alone implementations account for 16 of the 21 publications at the discrete decision-making level, where the model directly predicts whether and when a maneuver occurs. Examples include yellow-light decisions (Liu et al., 2024), emergency collision avoidance (Li et al., 2024), and decisions that involve gap acceptance such as left-turning (Bontje & Zgonnikov, 2024; Zgonnikov et al., 2024a), overtaking (Mohammad et al., 2023), and pedestrian crossing decisions (Giles et al., 2019; Pekkanen et al., 2022; Theisen et al., 2024; Tian et al., 2025). In these cases, the decision process can be represented directly by the canonical components of an EAM, as described in Section 2. A practical consideration in such tasks is determining the onset of the decision.

Traffic behavior often lacks a clearly observable point at which evidence accumulation begins, because relevant perceptual information, such as the approach of a vehicle, evolves continuously before the response. Decision onset therefore often has to be inferred from behavioral changes or environmental events, and different ways of defining it can lead to different conclusions about the underlying decision process represented by the model. One approach is the implementation of a discrete decision model embedded inside a larger framework that models continuous streams of evidence. In these applications, however, the EAM still only gathers evidence for a specific discrete decision (Engström et al., 2024; Markkula et al., 2016); in the braking models of Svärd et al. (2017, 2021) the accumulator also times the later brake adjustments of the same maneuver. In experimental settings, decision onset can sometimes be constrained more directly through design choices. In Zgonnikov et al. (2024a) the oncoming vehicle appears at the instant the driver stops at the line, so the moment at which evidence becomes available is set by the design; in Bontje and Zgonnikov (2024) the view of the intersection is obstructed until the decision can begin.

In contrast, five of the seven publications at the continuous action control level use embedded implementations. Here, accumulation processes operate within ongoing control loops. In longitudinal and lateral vehicle control, the EAM can function as an accumulator of control error that triggers a control adjustment once a boundary is reached. Examples include lane-keeping steering models (Markkula et al., 2018) and longitudinal control or car-following models (Durrani & Lee, 2024). Other publications embed EAM components within broader perception–action or interaction frameworks, modeling driver–pedestrian and gap-acceptance interactions (Markkula et al., 2023; Schumann et al., 2023) or collision avoidance (Schumann et al., 2026). In these contexts, the accumulation mechanism typically serves as one component within a larger perception–action architecture, that is, a model in which the accumulator is one stage in a loop that also represents how the scene is perceived and how the control action is generated, often replacing simpler threshold-based decision timing rules.

Table 2 also shows how these two dimensions relate. They are associated but not redundant: both implementations occur at both levels. The clearest illustration is within the continuous action control level itself: accumulation can be examined as a stand-alone account of when individual control adjustments are initiated, as in Durrani et al. (2021) and Goodridge et al.

(2023), or embedded within a broader control architecture that additionally specifies how the resulting action is generated and unfolds, as in Markkula et al. (2018) and Durrani and Lee (2024). The two dimensions therefore answer different questions: the modeling level describes the behavioral context, whereas the implementation describes how much of the surrounding perception–action process the model itself carries.

Overall, this categorization highlights that model implementation is related to, but distinct from, the behavioral context being modeled. The following section therefore examines how specific EAM components, such as drift dynamics, boundaries, and bias terms, are instantiated across these different modeling contexts.

## 4.3 Model Formulation: EAM Components and Parameterization

Nearly all reviewed publications use drift–diffusion models or their variants, that is, relative accumulators; only Bontje and Zgonnikov (2024) additionally evaluated a race model, and no application of the linear ballistic accumulator was identified, although the search query included it. The publications vary mainly in how three core components are specified: the drift, which governs how momentary evidence enters the accumulation process; the decision boundary, which specifies how much accumulated evidence is required for a response; and, in two-boundary models, the starting point, which can represent prior response bias. Columns 5–7 of Table 2 summarize these specifications.

### 4.3.1 Drift: Perceptual and Cognitive Evidence Accumulation

Across the reviewed publications, the drift reflects the evidence driving the change of the decision variable, and in 27 of the 28 publications at least one model links it to perceptual or kinematic information available to the road user. Some models instead estimate drift freely by condition (Liu et al., 2024; Theisen et al., 2024). This link partly follows from inclusion criterion 3, which requires the model to relate the behavior to the traffic situation. Unlike laboratory paradigms with constant within-trial evidence, traffic decisions often unfold under continuously changing conditions, motivating time- or state-dependent drift formulations. Direct model comparisons suggest that such formulations may help capture response-time patterns in pedestrian-crossing tasks, although their necessity and precise functional form remain task- and model-dependent (Ma et al., 2025; Theisen et al., 2024); in left-turn gap acceptance a time-varying drift alone did not capture the response-time pattern (Zgonnikov et al., 2024a). Noise in the accumulation process, by contrast, is treated uniformly: where it is included, it enters as an added diffusion term, and only one publication examines it beyond the standard assumptions, with a model variant in which cognitive load acts on the noise (Ma et al., 2026).

For example, in pedestrian-crossing tasks, the drift is, in at least one model of each of the six publications, a function of the approaching vehicle's time-to-arrival, sometimes combined with distance, speed, or rate of change (Giles et al., 2019; Ma et al., 2025, 2026; Pekkanen et al., 2022; Theisen et al., 2024; Tian et al., 2025). For drivers' braking, a drift depending on looming has been proposed (Markkula et al., 2016) and fitted to braking data (Durrani & Lee, 2023; Markkula, 2014; Svärd et al., 2021; Xue et al., 2018). Similarly, in left-turn gap acceptance and overtaking, the drift depends on a weighted combination of time-to-arrival and distance, the generalized gap (Bontje & Zgonnikov, 2024; Mohammad et al., 2023, 2024; Zgonnikov et al., 2024a, 2024b).

Beyond perceptual inputs, recent work examines how cognitive factors may influence drift dynamics. In particular, Ma et al. (2026) show that increased cognitive load reduces the responsiveness of the accumulation process, linking changes in drift to neural markers of evidence accumulation.

Across these applications, drift provides a mapping from perceptual or cognitive variables to momentary evidence. Its specification determines how changing traffic cues and, in some models, internal factors such as cognitive load enter the accumulation process.

### 4.3.2 Decision Boundaries: Fixed and Decreasing Formulations

The decision boundary is the level of accumulated evidence required for response commitment and is commonly interpreted as reflecting response caution (Bogacz et al., 2006; Forstmann et al., 2016). In time-critical tasks, the boundary may also vary over the course of a trial to represent increasing urgency, that is, growing pressure to act as the opportunity closes (Smith & Ratcliff, 2022). This is particularly relevant in traffic, where time pressure often changes as a situation unfolds and has motivated the use of decreasing boundaries.

Decreasing boundaries have been used in publications modeling pedestrian crossing (Ma et al., 2025, 2026; Theisen et al., 2024, in one of seven models), left-turn gap acceptance (Bontje & Zgonnikov, 2024; Zgonnikov et al., 2024a; tested in Zgonnikov et al., 2024b), overtaking (Mohammad et al., 2023; tested in Mohammad et al., 2024), and emergency collision avoidance, where the fitted boundaries hardly change (Huang et al., 2025; Li et al., 2024). In these publications, the boundary decreases within a trial because it depends on evolving traffic kinematics, such as time-to-arrival, gap, distance, or speed, rather than on elapsed time itself.

However, the empirical support for decreasing boundaries is mixed. Ma et al. (2025) report an improved fit for a decreasing boundary compared with a fixed one in a pedestrian-crossing task with a narrowing time window for action, and in left-turn gap acceptance only the model with a decreasing boundary captured the response-time pattern (Zgonnikov et al., 2024a). Other model comparisons find fixed boundaries performing competitively with decreasing ones (Mohammad et al., 2024; Zgonnikov et al., 2024b), and in Theisen et al. (2024) fixed boundaries estimated per condition fitted best on two of three criteria, whereas the decreasing boundary was best among the kinematics-dependent models. The clearest case is within a single research line: Mohammad et al. (2023) fitted only decreasing boundaries, whereas Mohammad et al. (2024) found that a decreasing boundary did not improve the fit and that a fixed boundary fitted the 2023 data about as well and preferred it as the simpler account. These comparisons suggest that the usefulness of decreasing boundaries depends on the task, the behavioral patterns being explained, and the other model components included.

Two publications that estimate boundaries separately across conditions, both modeling a two-choice decision, report a U-shaped pattern: the least evidence is required at intermediate values of the time-to-arrival (Liu et al., 2024; Theisen et al., 2024). The decreasing boundary functions used in the reviewed models cannot reproduce such a pattern.

Across the reviewed publications, the Decision boundary column of Table 2 records a boundary that is fixed within a trial in 17 cases and a decreasing boundary in at least one model in 10, including five in which only some models had one (Ma et al., 2025; Mohammad et al., 2024; Theisen et al., 2024; Zgonnikov et al., 2024a, 2024b), two of which preferred a fixed boundary (Mohammad et al., 2024; Zgonnikov et al., 2024b), while in Theisen et al. (2024) the preference depended on the criterion; the remaining publication fits no accumulator (Markkula et al., 2016). A boundary that is fixed within a trial can still differ between conditions, as in Liu et al. (2024) and in some models of Theisen et al. (2024). Decreasing boundaries are, however, not the only way to represent urgency: urgency-related gain or changes in drift can produce similar effects, so a better fit with a decreasing boundary does not by itself show which mechanism operates.

### 4.3.3 Starting Point and Non-Decision Time: Selective and Limited Use

The starting point may represent an initial response bias when it is displaced toward a particular decision boundary. In traffic applications, this parameter appears in fewer models than drift or boundary variations. Of the 28 publications, 12 model a decision between two identifiable alternatives, for which a starting point bias toward one of them is definable. Six of these fit a displaced starting point (Huang et al., 2025; Li et al., 2024; Mohammad et al., 2023, 2024; Theisen et al., 2024; Zgonnikov et al., 2024b); in a model of continuous control, Durrani and Lee (2024) allow asymmetric boundaries instead.

A starting point bias has a theoretical motivation when the road user has information favoring one response before the evidence of the trial accumulates; Mohammad et al. (2023), for example, introduced the driver's initial speed into the model components, including the starting point, because it is known to affect gap acceptance. Without such a reason, a better fit with a freely estimated starting point does not, by itself, demonstrate a prior response tendency. In Theisen et al. (2024), adding a starting point bias improved the fit of the condition-wise models, with a bias toward crossing when the approaching vehicle was initially farther away; in Mohammad et al. (2023, 2024), a starting point that depended on the driver's initial speed improved the fit for overtaking decisions. Starting point bias may therefore capture early response tendencies, but it is not a uniformly necessary component, and in traffic EAMs its role appears more limited and context-dependent than that of drift and boundary parameters.

Non-decision time captures sensory and motor delays outside the accumulation process, but its specification affects how the other parameters are interpreted. Fourteen publications include it. Eleven estimate it as a free parameter, several of them also modeling its variability (Bontje & Zgonnikov, 2024; Huang et al., 2025; Li et al., 2024; Liu et al., 2024; Ma et al., 2025, 2026; Mohammad et al., 2023, 2024; Theisen et al., 2024; Zgonnikov et al., 2024a, 2024b), and only Theisen et al. (2024) test a related choice, comparing models with and without variability in non-decision time. The other three model continuous action control and include perceptual and motor delays as separate parameters within the control loop, whereas stand-alone models can only identify their sum: Markkula et al. (2018) fix both delays, Goodridge et al. (2023) fix a motor latency, and Durrani and Lee (2024) estimate both within narrow bounds.

The drift, decision boundary, starting point, and non-decision time can trade off during estimation, which limits what model comparison alone can establish. Complexity varies widely, from publications that estimate no accumulator parameters to models with up to about twenty jointly optimized parameters.

## 4.4 Data, Model Fitting, and Model Evaluation

A further dimension of variation concerns the data used for model estimation, the fitting approach, and how the models were evaluated.

Regarding data, the reviewed publications draw on datasets that vary in size, experimental control and ecological validity, that is, how closely the recorded behavior resembles behavior in real traffic. Two main data sources can be distinguished: controlled experiments, mostly in driving or pedestrian simulators, and naturalistic driving or traffic-observation data. Controlled experiments are the most common, with sample sizes from 8 (Markkula et al., 2018) to 130 analyzed participants (Theisen et al., 2024, an online video experiment), and are used for discrete decision-making tasks such as pedestrian crossing, left-turn gap acceptance, braking, and overtaking (Giles et al., 2019; Ma et al., 2026; Mohammad et al., 2023; Pekkanen et al., 2022; Theisen et al., 2024; Zgonnikov et al., 2024a). Naturalistic driving and traffic-observation data are used or referenced as benchmarks in seven publications, for safety-critical braking events (Engström et al., 2024; Markkula et al., 2016; Svärd et al., 2017, 2021), car following (Durrani & Lee, 2024), road-user interactions (Schumann et al., 2023), and collision avoidance, with published naturalistic metrics as the benchmark (Schumann et al., 2026). The evidence rests on fewer independent sources than the number of publications suggests: the 28 publications draw on 26 distinct datasets, eight of which are analyzed in more than one publication (Supplementary Table S5). Li et al. (2024) and Huang et al. (2025), for example, report the same simulator experiment and share their rear-end parameter estimates. Model specifications are reused as well: ten publications use or extend closely related gap-based DDM specifications (lineage L03 in Supplementary Table S6).

Regarding model fitting, publications differ both in the estimation method and in the estimation level recorded in Table 2, that is, whether parameters were estimated per participant or for data pooled across participants. Estimation methods include likelihoods computed numerically from the model's response-time distribution (e.g., Theisen et al., 2024), likelihoods approximated from simulated responses (e.g., Giles et al., 2019; Svärd et al., 2021; Xue et al., 2018), least-squares or information-criterion-based objectives (e.g., Bontje & Zgonnikov, 2024; Mohammad et al., 2023), approximate Bayesian computation (Engström et al., 2024, on response times generated by their regression model), and grid searches or manual tuning of simulated models (Markkula et al., 2023; Schumann et al., 2026). Fourteen publications use pooled estimation, with one parameter set estimated from data aggregated across participants (e.g., Giles et al., 2019; Li et al., 2024; Xue et al., 2018), and Tian et al. (2025) adopt a previously estimated pooled parameter set without refitting it. Six publications report at least some participant- or drive-level fits (Durrani & Lee, 2023, 2024; Liu et al., 2024; Ma et al., 2025, 2026; Zgonnikov et al., 2024a), and Markkula et al. (2018) combine participant-specific and group-level estimation. Six publications do not estimate accumulator parameters from the analyzed data (Durrani et al., 2021; Engström et al., 2024; Goodridge et al., 2023; Markkula et al., 2016, 2023; Svärd et al., 2017). Only Theisen et al. (2024) additionally report a hierarchical Bayesian analysis, in which participant-specific parameters are estimated as draws from a common group-level distribution, to assess whether pooling affected the estimated parameter patterns. It reproduced the patterns of the pooled fits.

Regarding model evaluation, 23 publications compare the accumulation model with at least one alternative within the same publication, two of them qualitatively, without fitting (Durrani et al., 2021; Goodridge et al., 2023): 15 compare alternative accumulator specifications, nine compare it with a model of another kind and two remove components one at a time, some doing more than one. Table 2 gives the coding for every publication. The comparison criteria vary substantially, including likelihood-based measures and information criteria (e.g., the Akaike and Bayesian information criteria, AIC and BIC), prediction-error and distributional measures, and qualitative comparison of predicted and observed behavioral patterns. Evaluation on data not used in fitting, which can show whether a model generalizes beyond the data it was fitted to, is reported less often: four publications predict observations withheld from fitting (Durrani & Lee, 2023, 2024; Schumann et al., 2023; Xue et al., 2018), three apply a fitted or otherwise fixed accumulation model to an independent dataset (Markkula et al., 2023; Pekkanen et al., 2022, with two parameters re-estimated; Schumann et al., 2026), and four test predictions for an experimental condition withheld from fitting or tuning (Liu et al., 2024; Schumann et al., 2026; Zgonnikov et al., 2024a, 2024b); together, ten publications report at least one of these tests. Three publications report correspondence between an estimated quantity and a measure not used in fitting it: two with neural signals (Ma et al., 2025, 2026) and one with confidence reports (Bontje & Zgonnikov, 2024). However, these forms of evidence beyond model selection remain limited. Evaluations of a model other than the accumulation model (Engström et al., 2024; Markkula et al., 2016) are named in Table 2 but not counted here.

# 5. Synthesis of Findings

Section 4 described how EAMs are implemented and evaluated in traffic research. This section synthesizes what the reviewed publications show across studies. Overall, the reviewed work supports accumulation as an explanation of response timing, but accumulation models do not always predict better than other kinds of models. Most models let the drift change as the situation develops, whereas whether the boundary should also change over time differs between studies, and the models are useful for testing how road users process decision-relevant information. In more detail, four main findings emerge: the first two concern the support for accumulation and how accumulation models compare with other kinds of models, the third concerns which model components appear useful across different traffic tasks, and the fourth what accumulation models add to an analysis of traffic behavior.

*Accumulation versus fixed thresholds. First*, the reviewed comparisons provide support for integrating evidence over time rather than responding to momentary evidence alone. Of seven publications that compare or contrast an accumulation mechanism with a fixed perceptual threshold (Table 2, Models compared), five report findings more consistent with accumulation (Durrani & Lee, 2023; Durrani et al., 2021; Goodridge et al., 2023; Markkula et al., 2018; Xue et al., 2018), one finds a partial improvement (Durrani & Lee, 2024), and one favors the threshold model (Tian et al., 2025). These are not equally direct tests: three of the five fit both an accumulation and a threshold model to the same data (Durrani & Lee, 2023; Markkula et al., 2018; Xue et al., 2018), whereas two compare predicted and observed response patterns qualitatively, without fitting (Durrani et al., 2021; Goodridge et al., 2023). Nor are they independent: two analyze the same dataset (Durrani & Lee, 2023; Durrani et al., 2021), and the other three share an author who is also an author of this review (Goodridge et al., 2023; Markkula et al., 2018; Xue et al., 2018). Because each publication uses its own data and performance measures, the results show which model is favored within a study, but not how much better accumulation performs across publications. Observations of real-world emergencies are also consistent with integrating evidence over time. In naturalistic rear-end crashes and near-crashes, drivers' response times depended on how quickly the situation became critical, through the build-up of looming (Markkula et al., 2016) or of surprising evidence (Engström et al., 2024), rather than on a fixed delay. Accumulation predicts this pattern, but neither publication fitted an accumulation model to these observations, so the support is indirect.

*Comparison with other kinds of models*. Second, how accumulation models compare with other kinds of models depends on what is evaluated. Whereas the first finding concerns accumulation as a mechanism for response timing, this finding concerns how well the models as a whole predict behavior. Accumulation models predicted trajectories better than data-driven models (Schumann et al., 2023), reproduced the shape of car-following profiles better than established car-following models (Durrani & Lee, 2024), and avoided more simulated collisions than conventional driver models (Huang et al., 2025). On other outcomes in the same publications, however, they did no better: decisions (Schumann et al., 2023), pointwise car-following errors (Durrani & Lee, 2024) and braking-versus-steering choices (Huang et al., 2025). Elsewhere, a multilevel regression predicted brake timing better (Xue et al., 2018), and an accumulation model reproduced crossing decisions about as well as a neural network (Tian et al., 2025). Because the models were often calibrated differently or included accumulation as one component of a larger model, these comparisons do not show a general advantage or disadvantage of accumulation; whether an accumulation model is the better choice has to be evaluated for each application.

*Time-varying drift and boundary.* Third, most reviewed models let the drift follow the developing situation (Section 4.3.1). The few direct comparisons tend to support this, but not consistently. A time-varying drift fitted better than a constant one in pedestrian crossing (Ma et al., 2025) and among the models that link the drift to the traffic situation (Theisen et al., 2024), whereas in left-turn gap acceptance it captured the response-time pattern only together with a decreasing boundary (Zgonnikov et al., 2024a). Whether the boundary must also decrease differs between studies of similar tasks (Section 4.3.2). Because these studies use different comparison criteria and estimation levels, and the components trade off during estimation, the results do not yet show which components best account for a given task. The fitted models do, however, describe road users' decisions in a similar way. The evidence for going increased with the time-to-arrival of the approaching vehicle, distance added to time-to-arrival, and the starting point mostly favored committing to the maneuver (Mohammad et al., 2023, 2024; Theisen et al., 2024; Zgonnikov et al., 2024b). This agreement is partly built into the models and several of them are closely related, so it mainly shows that the models agree on how to describe the data.

*Testing how information is processed.* Fourth, accumulation models can be used to test hypotheses about how decision-relevant information is processed. Comparing model variants allowed publications to draw conclusions that a description of the behavior alone does not provide. For example, cognitive load was best modeled as a weaker response to the evidence, which EEG measures supported (Ma et al., 2026), and choices and response times did not distinguish between two accumulator architectures, whereas confidence ratings favored a single accumulator that kept accumulating after the decision (Bontje & Zgonnikov, 2024). Other publications used model comparisons to test which cues enter the evidence (Pekkanen et al., 2022) and how off-road glances change accumulation (Svärd et al., 2021), although each of these conclusions currently rests on a single publication.

All four findings rest on fewer independent sources than the number of publications suggests, because several publications share datasets or model specifications (Section 4.4).

# 6. Discussion and Outlook

This section discusses what the four findings of Section 5 imply for theory and future research (Section 6.1), for extended modeling approaches and their evaluation (Section 6.2), for methodological challenges (Section 6.3) and for practical applications (Section 6.4), followed by a concluding perspective (Section 6.5).

## 6.1 Theoretical Significance and Future Research

Traffic environments offer a valuable setting for cognitive modeling because they combine perceptual uncertainty, time pressure, and continuous perception–action coupling. The four findings of Section 5 point to several directions for future work.

First, the findings on response timing and on model components rest on evidence that changes during the decision, a condition that many traditional laboratory paradigms do not capture. Traffic applications therefore highlight opportunities to develop laboratory paradigms that better capture dynamic decision-making. Experiments involving sustained, time-varying sensory evidence (e.g., Winkel et al., 2014) and interactions between decision makers could help bridge laboratory and traffic EAM research. For example, U-shaped boundary estimates across time-to-arrival conditions (Liu et al., 2024; Theisen et al., 2024) motivate designs that systematically vary traffic conditions within participants. Another promising direction is the integration of behavioral and neurophysiological measures (Mulder et al., 2014; Turner et al., 2017). Recent pedestrian-crossing studies illustrate how EEG data can complement behavioral modeling of traffic decisions (Ma et al., 2025, 2026).

Second, the third finding indicates that the components favored by the available model comparisons differ between studies; whether accumulation parameters also differ between road users remains insufficiently examined, because many publications that estimate parameters pool data across participants (Section 6.3). Parameters such as drift, boundary separation, starting point, and non-decision time capture distinct aspects of decision-making (Ratcliff & McKoon, 2008; Ratcliff et al., 2016). Recent work incorporating participant-level random effects provides a promising example of how such variability can be modeled (Bachmann & van Maanen, 2024).

Third, building on the fourth finding that accumulation models provide testable accounts of information processing, traffic models could explore accumulation architectures beyond the single relative accumulator described in Section 2.1, including multi-dimensional or interacting accumulators, that is, models with more than one accumulator, whether these race independently toward separate boundaries or inhibit one another. Although such models exist in the broader EAM literature (Trueblood et al., 2014; Usher & McClelland, 2001), nearly all reviewed publications use a single relative accumulator, whether accumulating toward two boundaries or toward one; only Bontje and Zgonnikov (2024) additionally compared a race model, finding that confidence ratings distinguished the two architectures when choices and response times did not.

## 6.2 Embedded and Extended Modeling Approaches

Several reviewed publications embed accumulation in broader frameworks that also represent perception, interaction and control, such as surprise-based models of response timing (Engström et al., 2024), risk-sensitive models of emergency maneuvers (Huang et al., 2025), and models of road-user interaction that combine perception, inferences about the other road user's intentions (theory of mind), valuation of the possible actions and accumulation (Markkula et al., 2023; Schumann et al., 2023). Such integration allows EAM components to be used in models of interactions between road users, in traffic and potentially in laboratory paradigms, but it also makes it harder to identify which mechanisms drive behavior (Ratcliff et al., 2016; Smith & Ratcliff, 2022), which the two reviewed ablation analyses addressed by removing accumulation and other mechanisms one at a time (Markkula et al., 2023; Schumann et al., 2026).

## 6.3 Methodological Challenges

Several methodological challenges remain. Most traffic EAMs rely primarily on kinematic or visual input variables (e.g., Markkula et al., 2016; Zgonnikov et al., 2024a), although real-world decisions likely integrate multiple information sources, social cues, and multisensory inputs. While the importance of such integration is increasingly recognized, explicit computational modeling of multimodal integration remains relatively limited, as illustrated by related driving work on visual–vestibular cue integration outside the included set (Markkula et al., 2019).

Moreover, inter-individual variability receives limited attention: 14 publications fit one parameter set to data pooled across participants, and only one additionally reports a hierarchical analysis (Section 4.4). A distribution pooled across participants need not have the form of any individual participant's distribution, so parameters estimated from it need not describe the decision process of any individual, whereas both the modeled decision and the applied questions concern individual road users.

Interpretation of model parameters requires particular caution. When several parameters are allowed to vary simultaneously, including drift dynamics and decision boundaries, reliable attribution of behavioral effects to one parameter or another may become difficult, especially when the number of observations is low or the data are noisy (Bachmann & van Maanen, 2024; Dutilh et al., 2019; Ratcliff et al., 2016; Smith & Ratcliff, 2004). More broadly, parameter identifiability is a recognized challenge in computational modeling (van Maanen & Miletić, 2021), particularly as model complexity increases. These issues are compounded in naturalistic datasets where repeated observations per individual are limited. Recent methodological work proposes potential solutions (Bachmann & van Maanen, 2024), but their application in traffic modeling remains limited.

These considerations also constrain how far current traffic EAMs can be extended. Model fitting is already computationally demanding, often relying on global optimization or simulation-based methods, and additional mechanisms such as time-varying drift, decreasing boundaries, or embedded implementations can increase computational demands and complicate parameter interpretation (Engström et al., 2024; Theisen et al., 2024). Extensions are therefore most informative when they target a specific limitation of existing models and can be tested with data that distinguish competing mechanisms. The mixed evidence for brake-light information provides one example. In closely related braking models, the estimated contribution of brake-light information ranges from a large term that improves the fit only slightly (Xue et al., 2018) to none (Durrani & Lee, 2023), with mixed results in a third model (Durrani & Lee, 2024). Because brake-light onset and looming often change together (Durrani et al., 2021), their contributions would be better isolated in experiments that decorrelate the two cues. Such extensions are best added one at a time, each tested on data not used in fitting. Finally, identifying decision onset in continuous traffic behavior remains challenging (Markkula et al., 2018). As Section 4.2 describes, onset has to be inferred from behavioral changes, environmental events, or experimentally imposed markers. Different operationalizations of it may affect estimated response times and model parameters, and therefore the modeling conclusions drawn.

## 6.4 Practical Applications

Despite these challenges, the reviewed work points to at least one concrete practical modeling use. Evidence accumulation models can replace fixed reaction-time or perceptual-threshold rules in simulation of road-user behavior, such as microscopic traffic simulation and virtual safety testing, generating response timing from evolving evidence rather than from a fixed delay. This applies at both modeling levels: to discrete decision-making, such as at intersections and crossings, and to continuous action control, such as car following and steering.

Accumulation-based timing can reproduce some observed features of human traffic behavior in simulation, because responses depend on evidence that evolves over time. In car following, for example, replacing a zero or fixed reaction time with accumulation-based timing of control adjustments reproduced the intermittent adjustments of real drivers (Durrani & Lee, 2024). For applied models, this suggests a recommendation: where a simulation needs response timing, an accumulator driven by evolving kinematics, such as looming or time-to-arrival, is a promising alternative to a fixed reaction time (Section 5, first finding), whereas whether it also improves on other kinds of models has to be checked for each application (second finding).

Accumulation-based timing is also relevant to behavioral reference models, which serve as benchmarks for human behavior in the evaluation of automated driving systems. In such models, the timing of the response is important, because it determines how quickly an automated system is expected to respond. For instance, a reference model with a fixed reaction time was found to initiate braking later than human drivers did in real near-miss scenarios, whereas an accumulation-based model makes the response dependent on the rate at which a conflict situation develops (Engström et al., 2024; Olleja et al., 2025). Such accumulation-based models are already used in industrial practice (Engström et al., 2024; Schumann et al., 2026), for example as a reference model of human collision avoidance against which an automated driving system is assessed (Waymo, 2026).

More broadly, traffic EAM research may provide useful concepts for other applied domains in which action timing depends on evolving evidence under uncertainty, such as air-traffic control, maritime surveillance, healthcare, and other human-factors settings. This would build on existing work applying EAMs to safety-critical domains beyond traffic (Boag et al., 2023).

## 6.5 Concluding Perspective

Overall, traffic applications show that evidence accumulation models can account for a range of dynamic road-user behaviors. The synthesis of the reviewed publications provides relatively consistent support for accumulation-based response timing, although evidence that EAMs model other aspects of driving, such as decisions and full trajectories, better than other kinds of models remains mixed (Section 5). What EAMs do add is a testable account of how decision-relevant information is processed (fourth finding). This evidence comes mainly from simulator and laboratory experiments, with naturalistic driving or traffic-observation data used in seven of the 28 publications, and from a set of publications shaped by our choice to focus the search on applied, traffic-related work on driver decisions. This choice means that relevant work published in psychology venues may be missing (Section 3).

This review covered a subset of the literature in which EAMs were used to model decisions and control actions that road users use to navigate traffic. Related work outside our scope applies EAMs to driving as a context rather than as the decision of interest. Ratcliff and colleagues applied DDMs to simple driving tasks in a simulator to show how factors such as phone conversation, age and speed stress affect the drift, the decision boundary and the non-decision time (Ratcliff, 2015; Ratcliff & Strayer, 2014; Ratcliff & Vanunu, 2022; Vanunu & Ratcliff, 2023). In addition, studies of driver distraction modeled responses to a detection signal while participants were driving and having a conversation (Castro et al., 2019, 2023; Tillman et al., 2017). More recently, DDMs have been applied to the decisions of drivers in automated vehicles, for instance whether to take over control or to yield (Bao et al., 2026; Nie et al., 2026). Although these studies fall outside our scope, they show how factors such as distraction, age, workload and vehicle automation shape the decision processes that EAMs capture. Because they vary considerably in tasks, paradigms and modeling choices, synthesizing them with the work reviewed here is a natural next step.

For cognitive scientists, traffic offers a testbed for how evidence from changing environments shapes decisions, because it combines conditions that conventional laboratory tasks rarely address together: evidence that changes during the decision, time pressure, decisions about when rather than which, and interaction with other road users, in both discrete decisions and continuous control. In 27 of the 28 reviewed publications, at least one model links drift to perceptual or kinematic information from the traffic situation, such as looming, time-to-arrival, a gap, or a control error; many thereby represent changes in evidence within a decision. Traffic tasks thus offer a way to test how the evidence that drives a decision is derived from a changing environment, particularly when designs sample a wide range of conditions within participants and estimate parameters per participant or hierarchically rather than from pooled data, as in most publications that estimate parameters (Section 4.4). These conditions can also inspire new laboratory paradigms (Section 6.1).

For applied modelers, the same framework has been adapted to a wide range of road-user decisions and control tasks by changing the evidence input, the boundary, the starting point and the non-decision time. Which components improve an account of a particular task, however, has to be evaluated for that task: a decreasing boundary, for example, improved the fit in some studies but not in others. Ten of the 28 publications tested an accumulation model on data not used in fitting (Section 4.4). Models intended for application should therefore be tested on conditions or data not used in fitting, ideally in designs that separate correlated cues such as looming and brake-light onset (Section 6.3).

# 7. Declaration Statement

## 7.1 Competing Interests

The authors have no relevant financial or non-financial interests to disclose.

## 7.2 Data, Materials and/or Code Availability

No new empirical datasets were generated or analyzed during the current study. The literature reviewed in this article consists of previously published studies cited in the manuscript. Materials supporting the review methodology, including the exploration-phase procedure, guiding questions, keywords, search queries, the search and screening flow, and the publications excluded after full-text assessment, are available in the Supplementary Materials. No custom code was used in this study.